# One phase can overturn a stability verdict: reference-set dependence of thermodynamic stability in Mg–Ir–H, tested against the experimental record

**Authors.** Attia A. Gadallah[1], Sanaa Ismail[2*]

[1]Independent Researcher, State College, 16801 PA, USA

[2] Department of Materials Science and Engineering, The Pennsylvania State University, University Park, PA 16802, USA

***Corresponding author.** Sanaa Ismail sii5085@psu.edu

## Abstract

The energy above the convex hull is reported as a property of a compound. It is a property of the compound, and the phase set it was compared against, which the number does not show. We make that dependence quantitative: three studies disagree on whether cubic $Mg_2IrH_6$, a predicted ambient-pressure superconductor, is thermodynamically stable. Rebuilding the Mg–Ir–H hull from 1342 entries, only four of 212 absent compositions flip the sign alone; the study reporting stability held two of them on its own hull and names the two largest nowhere. The asymmetry has a structural reason: one added phase overturns a stable verdict, no single removal an unstable one. Against the laboratory record, the four synthesised hydrides the database holds lie within 9.1 meV/atom of the hull and three within 0.4, the compound never made is seven times further, a hydride made in 2002 is absent from all three databases holding this chemistry, and the compounds the data cannot represent at their own composition are the larger ones, on two held-out cases we name. Recomputing the ten decisive phases against pre-registered numbers returns 89.75 against the database's 85.49 meV/atom; three machine-learning potentials place the compound on their own hull.

## 1. Introduction

A stability verdict obtained from a convex hull is a comparative statement. The quantity reported as *energy above the hull*, $E_{hull}$, is the energy of a compound minus the energy of the cheapest linear combination of other phases with the same overall composition. Every term in that comparison except the first belongs to the reference set. Change the reference set and the number changes, with no indication in the number itself that anything has changed. The failure mode is therefore silent: a plausible value with no error flag.

That this can happen is not in doubt and has been discussed in general terms, including a recent reassessment that rebuilt complete phase diagrams for 603 reported stable structures and proposed chemical-space completeness as a confidence criterion [1]. Databases also provide quantities designed to expose it, such as the AFLOW pseudo-hull distance, the distance of a stable phase from a hull built without it [2]. What is missing is a worked case in which the dependence is measured against a published disagreement, attributed to named phases, and then checked against something other than another calculation.

$Mg_2IrH_6$ provides such a case, and it is consequential. A machine-learning-accelerated search over more than a million candidates proposed $Mg_2XH_6$ (X = Rh, Ir, Pd, Pt) as ambient-pressure conventional superconductors and placed the Rh, Ir and Pt members on the convex hull of their ternary systems [1]. A second study put the same compound about 60 meV/atom above its hull and proposed a synthesis route through an insulating $Mg_2IrH_7$ intermediate [2]. A third synthesised $Mg_2IrH_5$, isostructural but with 5/6 hydrogen site occupancy, found it the more stable of the two to 28 GPa, and argued that configurational entropy favours it [3]. Experiment has since produced cubic $Mg_2IrH_7$ above 40 GPa and cubic $Mg_2IrH_5$ on decompression, and reports that $Mg_2IrH_6$ does not form spontaneously, presenting it instead as a target for non-equilibrium processing [4]; the rhodium analogue has been made above 30 GPa, with a transition temperature reaching 29 K at 53 GPa [5]. Supplementary Note 1 gives the fuller survey.

The three computational verdicts on the hull, about 60 meV/atom above it, and outcompeted by the $H_5$ phase are not obviously reconcilable, and the literature notes the tension without resolving it. We take the position throughout that none of these studies is in error. Each is internally consistent with its own reference set and its own structure search. The decisive phase is named nowhere in the earliest of them, and the dates make that unsurprising rather than careless: the rhombohedral $Mg_2IrH_7$ was proposed in the second study, posted eight days after the first. The disagreement is a difference in an input that was not reported as an input. Our contribution is to make that input measurable, to measure it, to test the result against the laboratory record, and to recompute the phases that decide it under a setting of our own, against numbers registered before the calculation ran.

The paper is organized as a single argument. Section 3 reconstructs the hull and attributes the disagreement to individual phases. Section 4 shows why the hydrogen reference, the other candidate explanation, cannot account for it, for a reason that generalises. Section 5 validates the resulting hull against the experimental record of the system, and then turns the same test on the reference data themselves, measuring a cell-size ceiling across 29 systems and asking, against criteria fixed in advance, two questions it does not answer in our favour  whether the ceiling belongs to that database or to the reference data of the field, which is inconclusive, and whether a global bound on cell size demonstrably excludes measured phases outside the two systems that generated the hypothesis, which it does not  and one it does: whether the data represent each measured composition with a cell large enough to be that compound. Supplementary Note 6 reads the same energetics as hydrogen-release thermodynamics, which states the size of the disagreement in units a hydrogen-storage reader can act on. Section 7 recomputes the decisive phases ourselves, under one density-functional setting and against numbers registered before submission, so that the result does not depend on the source being internally consistent. Section 8 asks whether a fast universal potential could have decided the question, under criteria fixed before the calculation. Section 9 states what metadata a reported hull distance requires.

# 2. Methods

## 2.1 Discipline on the comparison

Every statement in this paper obeys three rules, enforced in code rather than by convention.

1. **No energy is compared across exchange–correlation functionals.** Phases carry their source as a field, and the phase-diagram constructor refuses to build a diagram from phases of mixed source. Comparisons between functionals appear only as comparisons of *verdicts*, never of energies.
2. **One settings string per calculation campaign.** Every structure relaxed with a machine-learning potential carries a settings string `MACE/medium/float64/2stage/disp=False/fmax=0.02` for the first gate and the analysis asserts that exactly one distinct string is present across all rows before computing anything. The density-functional campaign of Section 7 obeys the same rule by construction: one input template per stage for all ten phases, and the output files of all ten runs name the same three pseudopotentials and no others.
3. **Counterfactuals are of the phase set, not of the energies.** When we report the effect of adding or removing a composition, the energies of all remaining phases are unchanged and come from one source.

## 2.2 Data, and what is computed here

The two halves of this paper differ in kind. **Sections 3 to 6 recompute nothing** Supplementary Note 6, the hydrogen-release reading, is supplementary : they re-analyse published database energies, and what we compute from them hull energies, decomposition facets and the sensitivity of both to the phase set are geometric consequences of energies that already exist. **Sections 7 and 8 are our own calculations**: a density-functional campaign of 60 tasks and 20.6 core-hours on the ten phases that carry the result, and 93 structure relaxations repeated for each of three machine-learning potentials. The subset of the database energies that decides the verdict is independently recomputed in Section 7.

Reference energies and structures were obtained from the Alexandria database of density-functional calculations [9] (PBE and PBEsol functionals) and from the OQMD [10], through their OPTIMADE interfaces, in a snapshot taken 2026-09-15 and archived with the study. Snapshot contents for the Mg–Ir–H system:

The PBE snapshot holds 1342 entries over 241 compositions, 84 of them pure hydrogen and 53 of those molecular; the PBEsol snapshot holds 90 entries over 29 compositions and no molecular hydrogen phase at all; the OQMD snapshot holds 7 compositions. The counts are tabulated in Supplementary Note 2.2.

The asymmetry in the final column is a property of the data, not an artefact of our selection, and is used in Section 4. The survey of Section 5.4 queries the same Alexandria PBE endpoint for 29 ternary systems rather than one, and then repeats a single query one filter, one field against the Materials Project, the OQMD, AFLOW and NOMAD through the same OPTIMADE interface, so that what differs between providers is the data and not the question. Counts come from each provider's own `meta.data_returned` on a filtered query, which is exact without paginating; cell sizes are enumerated entry by entry where a provider does not serve a filter on `nsites`. The experimental inventory of Supplementary Note 5.5 is taken from a crystallographic database of measured structures [21], whose listing metadata composition, Pearson symbol and space group are public, and from which we use the Pearson symbol because it carries the number of atoms in the conventional cell. The hydrogen reference in these entries is a corrected value, 28 meV/atom above the lowest computed hydrogen phase; being uniform in hydrogen content it cannot move a hull distance, for the reason given in Section 4.2.

## 2.3 **Computational settings (DFT)**

The ten decisive phases cubic and C2/m $Mg_2IrH_6$, the two decomposition-facet phases ($Mg_2IrH_5$ and $Mg_2IrH_7$), the three nearest competing phases ($Mg_2IrH_4$, $Mg_3Ir_2H_{12}$, and $Mg_2IrH_3$), and the elemental references (Mg, Ir, and the $H_2$ molecule) were recomputed from the snapshot starting structures using the PBE functional without spin polarisation. Calculations used a 600 eV plane-wave cutoff, Γ-centred k-point meshes at 0.20 $Å^{-1}$ spacing, and the projector-augmented-wave potentials PAW_PBE Mg_pv (13Apr2007), PAW_PBE Ir (06Sep2000), and PAW_PBE H (15Jun2001). Each crystal was relaxed in two sequential stages (restarting the second from the first to rebuild the basis at the updated cell) to a force tolerance of 0.01 eV/Å and an electronic convergence of $10^{-6}$ eV using Gaussian smearing (0.05 eV), followed by a final static calculation converged to $10^{-7}$ eV with the tetrahedron method and Blöchl corrections. The isolated $H_2$ molecule was treated in a fixed cell at a single k-point with 0.01 eV Gaussian smearing. Convergence of the signed hull distance itself was verified across 30 static runs spanning 520–700 eV cutoffs and 0.15–0.25 $Å^{-1}$ k-point spacings, yielding a total spread of 0.42 meV/atom against the pre-registered tolerance of 5 meV/atom.

## 2.4 Signed distances and the exclusion rule

To report a *signed* distance for $Mg_2IrH_6$, the hull must be built without any $Mg_2IrH_6$ entry; otherwise, the compound is compared against itself and the distance is clipped at zero. All signed distances in this paper exclude every entry of the target composition. This is the quantity AFLOW calls the pseudo-hull distance [12]; we adopt it and the term.

### 2.5 Attribution protocol

Let **C_full** be the 241 compositions of the PBE snapshot, and **$C_{red}$** the 29 compositions present in the PBEsol snapshot; Fig. 1b shows both on the ternary triangle, and where the second one is thin. Using PBE energies throughout:

- **Baseline:** hull built from PBE entries restricted to **$C_{red}$**.
- **Add-one:** for each composition in **$C_{full}$** \ **$C_{red}$** (212 of them), add that composition's PBE entries to the baseline and recompute the signed distance.
- **Remove-one:** for each of the same compositions, remove it from the full set and recompute.
- **Greedy:** repeatedly add whichever composition raises the distance most, until the sign changes.

The add-one and remove-one tests answer different questions how fragile is "stable", and how robust is "unstable" and their asymmetry is the main result of Section 3.

**What $C_{red}$ is, and what it is not.** It is a smaller reference set that exists and can be enumerated, so it is a usable instance on which to *measure* phase-set dependence. It is **not** a reconstruction of any published study's reference set: of the nine phases reference [1] names on its own hull, exactly one the target itself is in **$C_{red}$**. Section 3.5 asks what that study held, from its own text.

## 3. The disagreement is attributable to individual phases

### 3.1 The two endpoints

With PBE energies throughout and the hull built without any $Mg_2IrH_6$ entry:

| Reference set | Signed distance of cubic $Mg_2IrH_6$ | Decomposition |
|---|---|---|
| All 241 compositions | **+85.5 meV/atom** | 0.444 $Mg_2IrH_5$ (I4mm) + 0.556 $Mg_2IrH_7$ (R-3m) |
| Restricted to the 29 compositions of the reduced set | **−63.6 meV/atom** | 0.556 $MgIrH_3$ + 0.333 $MgH_2$ + 0.111 H |

The coefficients are atom fractions, not formula units: in formula units the facet is half $Mg_2IrH_5$ and half $Mg_2IrH_7$, which is why hydrogen balances across it. The two numbers differ by 149 meV/atom and in sign (Fig. 1a).

The second row is a construction, so we checked the verdict without it. On its **own** energies the PBEsol snapshot places cubic $Mg_2IrH_6$ 162.9 meV/atom **below** its hull and holds no other ternary hydride there at all, while the PBE snapshot holds six ternary hydrides on its hull and $Mg_2IrH_6$ is not among them. That contrast is the paper's thesis in one sentence, and it removes the objection that the −63.6 is an artefact of borrowing energies across functionals. Comparing verdicts between functionals is permitted by the first rule of Section 2.1; comparing their energies is not, and we do not. Each hull was computed by two implementations sharing no code, which agree exactly (Supplementary Note 3.1).

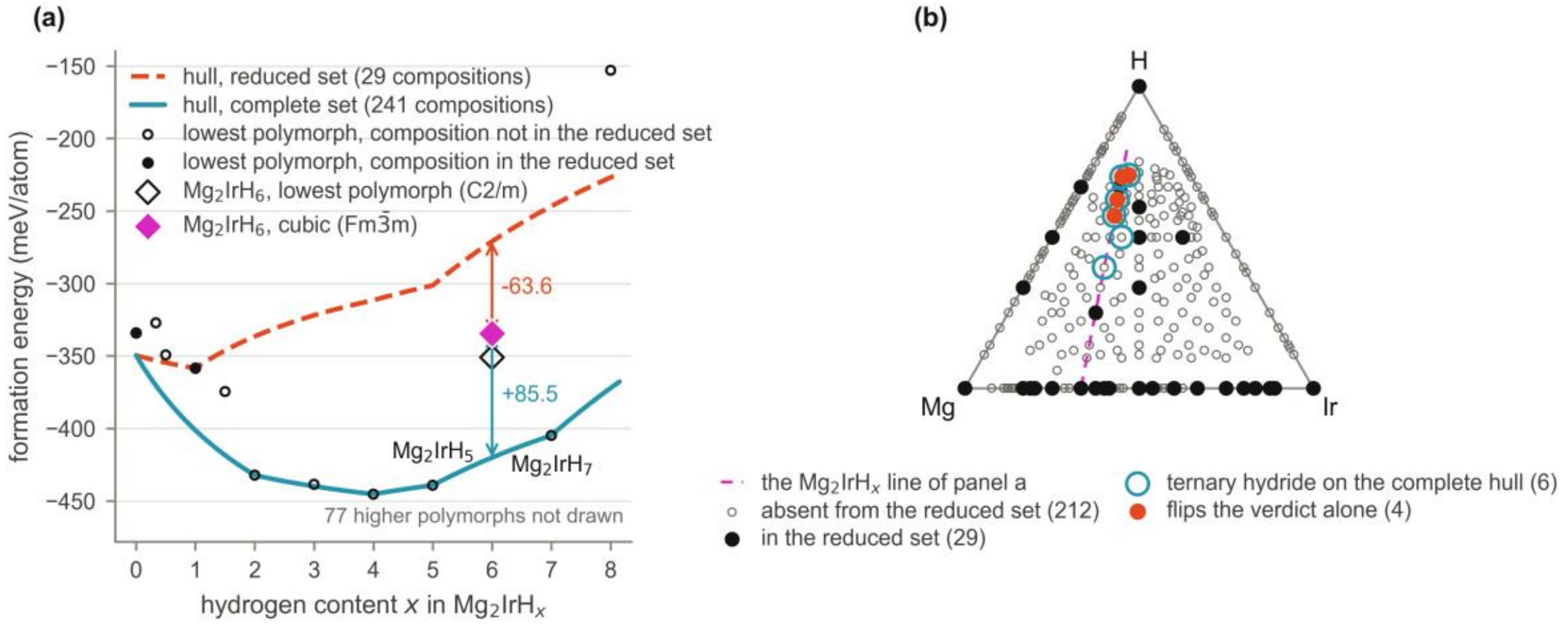

**Figure 1.** (a) Formation energy against hydrogen content along the $Mg_2IrH_x$ line. Both hull sections are evaluated exactly at every composition and built without any $Mg_2IrH_6$ entry, so the gaps shown are the signed distances quoted in the text: the complete 241-composition set places cubic $Mg_2IrH_6$ 85.5 meV/atom above the hull, the 29-composition reduced set 63.6 meV/atom below. They come from the same energies, the same functional and the same code, and differ only in which compositions were allowed to compete. (b) The same two reference sets drawn on the Mg–Ir–H triangle, one point per composition placed at its lowest entry. A scatter and not a triangulated hull, because the quantity here is coverage and a triangulation would be dominated by the complete set while the reduced set's holes the content would disappear between the tie lines. The reduced set holds the Mg–Ir edge and little of the interior; every one of the four compositions that flips the verdict alone, and five of the six ternary hydrides on the complete hull, lie in the hydrogen-rich region it does not cover.

## 3.2 Which phases carry the effect

Of the 212 compositions present in the full set and absent from the reduced one, nine lie on the PBE hull, and **four** of those nine, added one at a time to the baseline, move the signed distance above zero. They are $\mathbf{Mg_2IrH_7}$ (to +67.6 meV/atom, a swing of 131.2), $\mathbf{Mg_3Ir_2H_{12}}$ (+63.9), $\mathbf{Mg_2IrH_5}$ (+58.9) and $\mathbf{Mg_2IrH_4}$ (+18.0), all four labelled on Fig. 2b; Supplementary Note 3.2 adds the space group of each lowest entry.

The greedy search terminates after a single addition: no group of phases is required (Fig. 2b). The remaining 208 absent compositions, including every Mg–Ir intermetallic absent from the reduced set, do not change the sign. What that addition does to the hull is drawn in Fig. 2a: the new composition becomes a vertex, the facet beneath $Mg_2IrH_6$ re-forms around it, and the compound that was 63.6 meV/atom below the surface is left 67.6 above it, without one energy having changed. Coverage failure in this system is therefore not diffuse it is not a matter of database size but sharp: the verdict depends on a small number of chemically adjacent phases.

Those phases are not an arbitrary subset. $Mg_2IrH_5$ and the $Mg_3Ir_2H_{12}$ stoichiometry are charge-balanced arrangements of closed-shell eighteen-electron complex anions, whereas $Mg_2IrH_6$ requires a nineteen-electron $[IrH_6]^{4-}$ unit, one electron beyond the closed shell which is also why its band structure is metallic and why it is a superconductor candidate at all. That complexes of this class are commonly eighteen-electron and less often sixteen or fourteen is established in the reference literature [18], and that $Mg_2IrH_5$ in particular is charge-balanced is the argument of the study that made it [3]; we claim none of it. What the attribution adds is the connection: the compositions whose absence produces the stable verdict are precisely the closed-shell competitors of the open-shell metal under test.

## 3.3 The asymmetry, and which verdict is robust

Removal is not the mirror of addition. Starting from the complete set:

Removing $Mg_2IrH_5$ alone leaves +80.7 and removing $Mg_2IrH_7$ alone leaves +81.1; removing **both** members of the facet at once leaves **+74.8**; removing any other single composition leaves +85.5 unchanged (Supplementary Note 3.3).

No single removal, and not even the simultaneous removal of both decomposition products, restores stability (Fig. 2b).

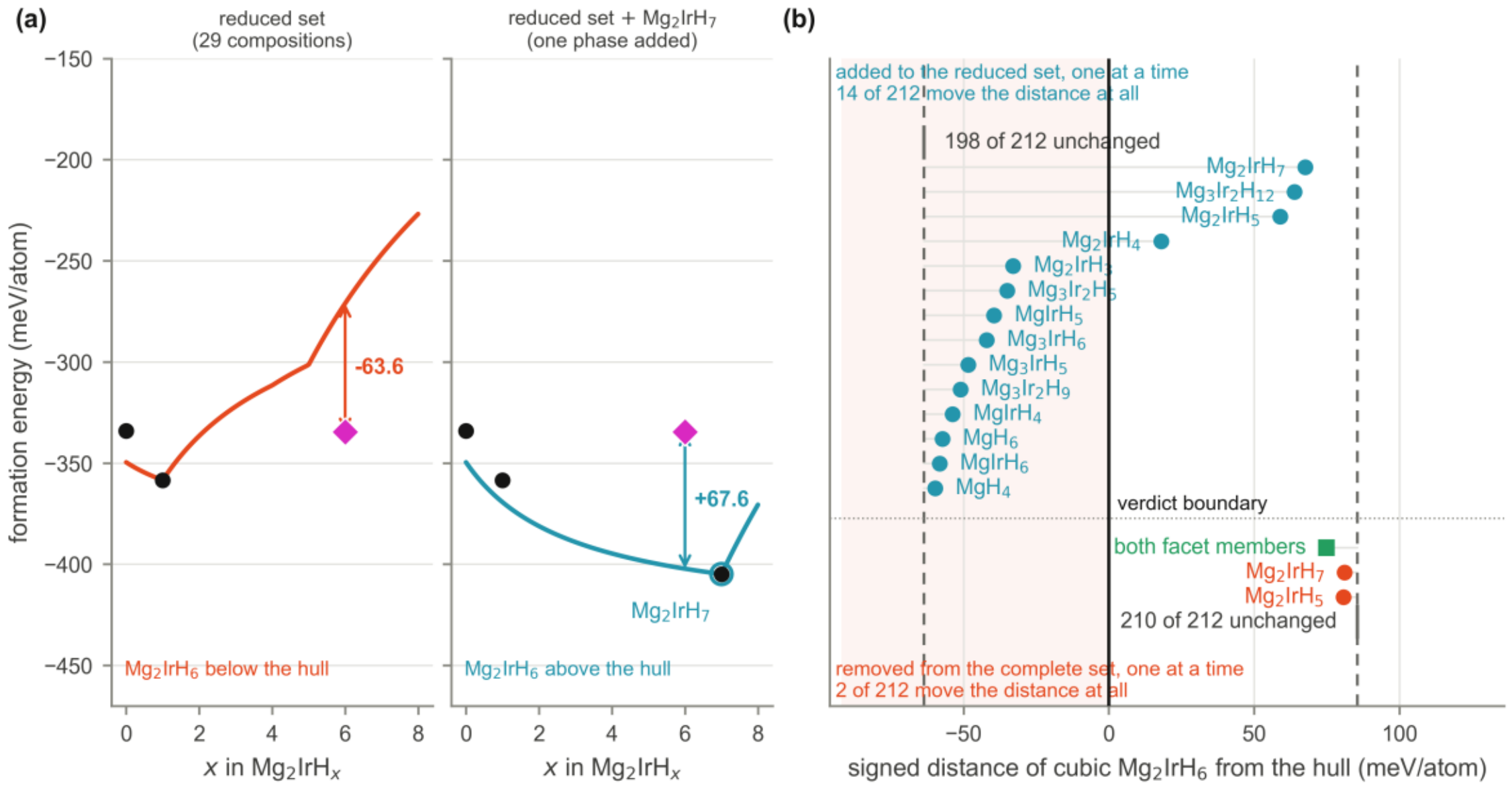

**Figure 2.** One phase overturns the verdict; no phase restores it. PBE energies throughout, and every hull built without any $Mg_2IrH_6$ entry so that each distance is signed. (a) The $Mg_2IrH_x$ section of the hull built from the reduced reference set (left) and from the same set

with $Mg_2IrH_7$ added (right), at identical scale. Adding the one composition introduces a vertex, re-forms the facet under the target, and moves cubic $Mg_2IrH_6$ from 63.6 meV/atom below the surface to 67.6 above it. (b) All 212 absent compositions added to the reduced set one at a time and removed from the complete set one at a time, on one axis. Only 14 and 2 of them respectively move the distance at all, so the rest are collapsed into one row each rather than drawn as 408 coincident points. Four additions cross the verdict boundary; no removal does, not even both members of the decomposition facet together. The shared axis is the point: a stable verdict from an incomplete set is overturned by one phase, an unstable verdict from a complete set is not.

This is a structural property of the hull rather than a coincidence: when a compound is above the hull, it is above it with respect to a facet supported by several phases, and deleting one vertex of that facet exposes the next one. The consequence for interpretation is direct and, we believe, the most useful single sentence in this section:

> **A "stable" verdict obtained from an incomplete reference set is fragile one phase can overturn it. An "unstable" verdict is robust no single phase can.**

### 3.4 How large an input error would have to be

A difference of 149 meV/atom is not by itself evidence of robustness; what matters is how well the energies entering the comparison are controlled. The question can be answered exactly, by perturbing the inputs and re-solving:

Raising every $Mg_2IrH_5$ entry, every $Mg_2IrH_7$ entry, or both together changes nothing at any magnitude up to 4 eV/atom; deleting both facet compositions outright moves the distance only to +74.8; deleting any other single composition changes nothing. The one perturbation that flips the verdict is a fall of 85.5 meV/atom in the target's own formation energy (Supplementary Note 3.4).

The verdict is therefore **immune to arbitrary error in the competing phases**: raising or deleting them exposes the next phase and the hull stays below the target. The whole of the remaining uncertainty is one number whether the target's own formation energy, on this database's footing, is 85.5 meV/atom too high.

### 3.5 What the published study held, from its own text

Sections 3.1 to 3.4 measure how much a hull distance depends on its phase set. They do not, by themselves, say that this is what happened in the published disagreement. That is a claim about a particular study, and it must be answered from that study, so we answer it from its text rather than from our baseline.

The machine-learning study [1] states that its Mg–Ir–H phase diagram was "computed with PBEsol" from experimental phases, a prototype search and global structure prediction, and names nine phases on it. The comparison that matters is where our own hull places each of those nine, and we report it in full because a partial comparison was how an earlier version of this section went wrong (Supplementary Note 3.5 quotes the passage verbatim):

| Reported stable by [1] | On our complete-set PBE hull |
|---|---|
| $Mg_3Ir_2H_5$, $Mg_2IrH_2$, $Mg_2IrH_4$, $Mg_2IrH_5$ | **on the hull** |
| $Mg_4IrH_5$ | 0.4 |
| $Mg_5Ir_2H_2$ | 4.8 |
| **$MgIrH_5$** | **48.0** |
| **$Mg_2IrH_6$** | **69.1** (lowest polymorph), 85.5 (cubic) |
| **$MgIrH_6$** | **77.3** |

and our hull carries two phases that list does not report: $Mg_2IrH_7$, and $Mg_3Ir_2H_{12}$, which is the most load-bearing phase on it removing it would let the hull fall 58.2 meV/atom at its own composition.

**So, the two hulls do not differ in one place.** Five of the nine agree, one more is within 5 meV/atom, and three are placed 48 meV/atom or further apart two of those by amounts comparable to the 85.5 this paper is about. $MgIrH_5$ and $MgIrH_6$ are iridium-rich compositions among the residuals of our own reproduction check (Supplementary Note 2.3), where the source reports 44.0 and 70.3 against our 48.0 and 77.3, so part of that gap is ours; what remains is still a hull that lands elsewhere across the hydrogen-rich region, and a difference of that shape is not reducible to one absent phase. The second study [2] places the rhombohedral $Mg_2IrH_7$ on its hull and $Mg_2IrH_6$ above it.

With that on the table, what can and cannot be said about the four compositions of Section 3.2:

| Composition whose addition flips the sign | Swing | Named by [1] |
|---|---|---|
| **$Mg_2IrH_7$** | **131.2** | **no, the formula occurs nowhere in that paper** |
| $Mg_3Ir_2H_{12}$ | 127.5 | not named, and not excluded: its Fig. 2 carries unlabelled metastable points |
| $Mg_2IrH_5$ | 122.5 | **yes, on its own hull** |
| $Mg_2IrH_4$ | 81.6 | **yes, on its own hull** |

Two of the four were in that study's reference set and on its hull, so their absence cannot be what produced its verdict. That leaves two candidates, $Mg_2IrH_7$ and $Mg_3Ir_2H_{12}$, whose swings differ by 3.7 meV/atom, and the ground for preferring the first that its formula occurs nowhere in that paper applies to the second as well, since neither is named. The timing is at least consistent for $Mg_2IrH_7$: it was proposed in the second study [2], posted eight days after [1].

**But the first table limits how far this can be pushed.** An attribution to one absent phase would require the two hulls to agree elsewhere, and they do not; a gap of 48 and 77 meV/atom on two of that study's own stable phases is the signature of a different functional and a different structure search. What the section establishes is therefore narrower than it first appears and still worth having: the published disagreement is **consistent with** a reference-set difference of the kind Sections 3.1 to 3.4 measure, two of the four candidates are ruled out by that study's own report, and the remaining two are absent from it but the reference set is one of at least two explanations of comparable size, and we have not separated them.

Three things this does not establish that [1] never computed the phase, that the functional plays no part, and that C_red reconstructs its reference set are set out in Supplementary Note 10 and in Supplementary Note 3.5.

What is measured, rather than inferred, is that **one phase can carry the whole of a verdict**: on one set of energies, adding $Mg_2IrH_7$ alone to the reduced baseline moves the signed distance by 131 meV/atom and changes its sign. $Mg_2IrH_7$ was subsequently synthesised [4].

---

# 4. The hydrogen reference cannot be the explanation

## 4.2 Why a composition-linear shift cannot move a hull distance

The two snapshots do disagree about hydrogen, and measurably: over 23 hydrides with 8 intermetallic controls, the difference between them is proportional to hydrogen content at **−0.561 ± 0.054 eV per hydrogen atom**, with a coefficient of determination of 0.956, and the PBEsol snapshot contains no molecular hydrogen phase at all. That is a real defect of one reference set, measured here for the first time to our knowledge and reported in full in Supplementary Note 4.1. It is also incapable of producing the disagreement, for the following reason.

Let the energy of every phase be shifted by a term that is linear in composition, $E \rightarrow E - \Sigma_e c_e x_e$, where $x_e$ are atomic fractions summing to one. Both the compound and every linear combination of phases at the same composition receive exactly the same shift, so the difference between them the hull distance is invariant. A uniform change in any elemental reference energy, including the hydrogen reference, is of this form.

We confirmed the invariance numerically twice, in two independent settings. Rebuilding the hull with the hydrogen reference displaced by ±100 meV/atom leaves the distance of $Mg_2IrH_6$ unchanged (Supplementary Note 8.3). Fitting the per-element error of a machine-learning potential and subtracting it from every phase moves the signed distance by 0.0000 meV/atom (Supplementary Note 8.5).

# 5. Testing the hull, and the reference data, against the experimental record

## 5.1 The test

Internal statistics cannot establish that a hull is right. For this system an external test is available, because the experimental record is small enough to be enumerated. We list it in full and ask where the computed hull places each compound. Two sources are needed and we name both, because neither contains the other: the recent literature, which carries the high-pressure phases of 2024 to 2026, and the crystallographic inventory of measured structures harvested for Supplementary Note 5.5, which carries older phases the literature search did not surface. The list is declared data with sources; if a row is wrong, the conclusion changes.

| Compound | Experimental status | In the PBE snapshot | Distance from the computed hull |
|---|---|---|---|
| $Mg_2IrH_5$ | synthesised at mild pressure–temperature conditions [3] | 9 entries, sg 107 | **0.0** |
| $Mg_2IrH_7$ | synthesised above ≈40 GPa [4] | 2 entries, sg 166 | **0.0** |
| $Mg_4IrH_5$ | known ternary hydride [18] | 6 entries, sg 8 | **0.4** |
| **$Mg_3IrH_5$** | **measured structure, mS36, sg 12** [21] | 2 entries, sg 5 | **9.1** |
| $Mg_6Ir_2H_{11}$ | **synthesised 2002, $P2_1/c$** [7] | **absent** | |
| $Mg_3Ir$ | intermetallic, hydrogenation precursor [7] | 16 entries, sg 185 | 0.0 |
| MgIr | intermetallic | 37 entries, sg 64 | 0.0 |
| $Mg_2Ir$ | intermetallic | 19 entries, sg 63 | 15.5 |
| $Mg_5Ir_2$ | decomposition product [7] | 1 entry, sg 72 | 25.6 |
| **$Mg_2IrH_6$** | **not synthesised; presented as a target for non-equilibrium processing** [3, 4] | 7 entries | **69.1** (lowest polymorph), **85.5** (cubic) |

Five hydrides of this system have been made, and the database contains **four** of them; $Mg_6Ir_2H_{11}$ is absent. Of those four, **three lie on the computed hull or within 0.4 meV/atom of it and the fourth, $Mg_3IrH_5$, lies 9.1 meV/atom above** (Fig. 3). **Every one of the four is at least 7.6 times closer to the hull than the lowest polymorph of the compound that has never been made.** The ordering, not a threshold, is what the test delivers: the made compounds are near the hull and the unmade one is far from it. The comparison is among hydrides; the two made intermetallics of the same table sit at 15.5 and 25.6 meV/atom, which narrows the factor to 2.7 if they are included, and we state both rather than the favourable one.

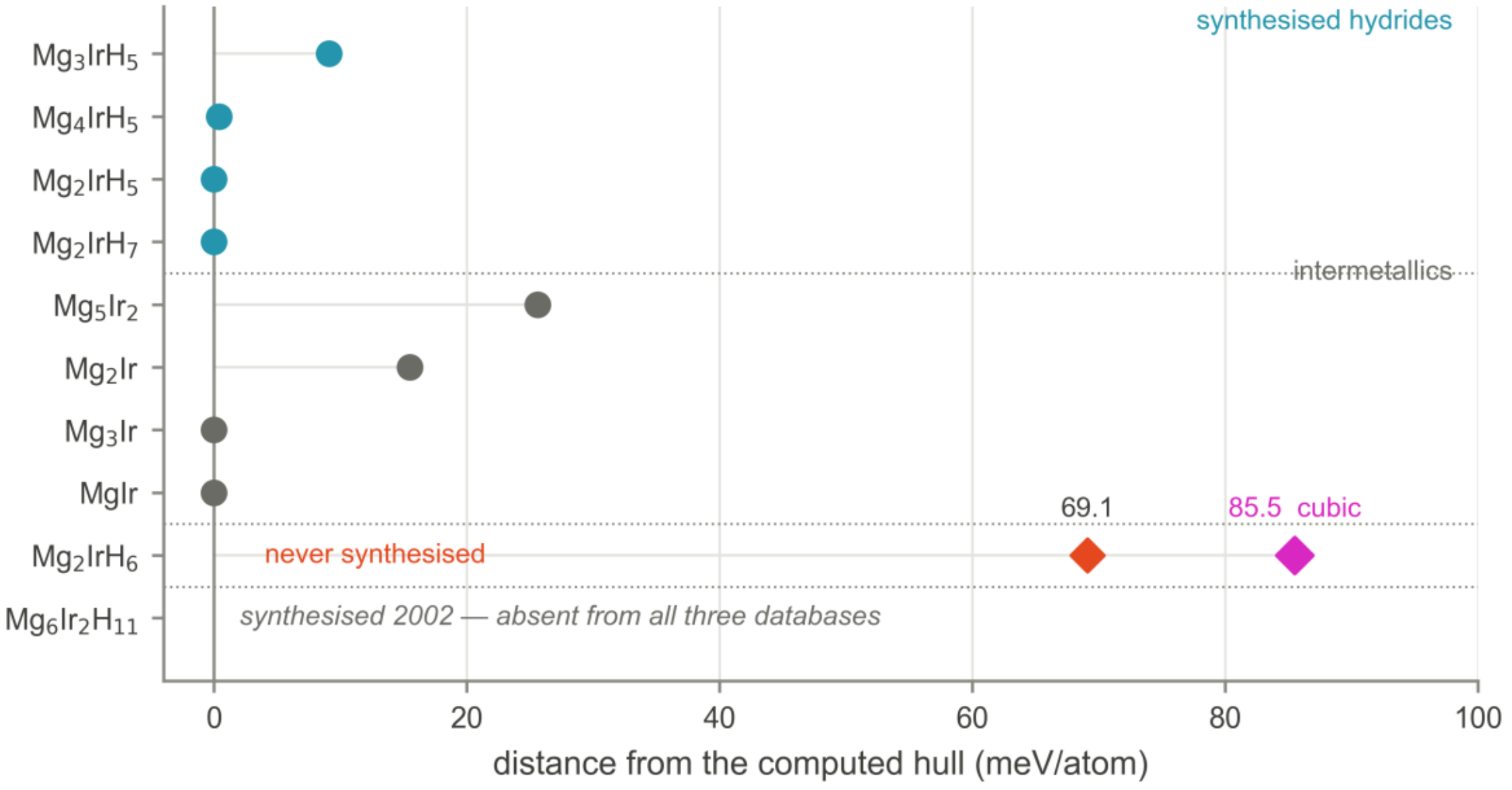

**Figure 3.** Distance from the computed hull for every compound of this system whose experimental status is known, sorted within each group. The four synthesized hydrides the database holds lie between 0.0 and 9.1 meV/atom of the hull and three of them within 0.4, while the compound never synthesised is at least 7.6 times further from it than any of them. The two values shown for $Mg_2IrH_6$ are its lowest polymorph and its cubic form. $Mg_3IrH_5$, the one made hydride not within rounding of the hull, is also the one whose stored cell is half the size of its measured structure (Section 5.6), so its 9.1 is an upper bound. $Mg_6Ir_2H_{11}$, synthesised in 2002, has no entry in any of the three databases queried and is shown without a marker.

The row not within rounding of zero is also the one Section 5.6 explains: the database holds nine atoms at the $Mg_3IrH_5$ composition where the measured structure needs eighteen, so its stored energy is too high, and its 9.1 meV/atom is an **upper** bound. $Mg_4IrH_5$ carries the same caveat ten atoms stored against twenty measured so its 0.4 is an upper bound too, which does not change the ordering.

## 5.2 The objection this forecloses

A complete hull built from a computational database necessarily contains predicted phases that have never been made, and a referee is right to ask whether the verdict depends on them. Of the six ternary hydrides on our PBE hull, only two $Mg_2IrH_5$ and $Mg_2IrH_7$ are experimentally known; the other four ($Mg_3Ir_2H_{12}$, $Mg_2IrH_2$, $Mg_2IrH_4$, $Mg_3Ir_2H_5$) are predictions. But the decomposition facet of $Mg_2IrH_6$ is 0.444 $Mg_2IrH_5$ + 0.556 $Mg_2IrH_7$ in atom fractions, half of each in formula units, and **both of those compositions have been** **synthesised.** The verdict rests on made compositions and requires none of the four predicted hull phases.

That defence is at the level of composition, and this paper is about the difference between a composition and a structure, so we apply our own distinction to it: **neither facet member is, as a structure, the one experiment made** R-3m $Mg_2IrH_7$ at zero pressure against a cubic phase made above 40 GPa that reverts below 20, and an ordered I4mm $Mg_2IrH_5$ against a disordered 5/6-occupancy one. The composition-level statement stands; whether R-3m $Mg_2IrH_7$ is a legitimate zero-pressure competitor is open, it would take phonons on a 10-atom cell to settle, and it is the sharpest unaddressed objection to the central number (Supplementary Note 5.2).

## 5.3 A compound made in 2002 is missing from three public databases

$Mg_6Ir_2H_{11}$ was synthesised by hydrogenation of the intermetallic $Mg_3Ir$ at 20 bar and about 300 °C, and alternatively by sintering the elements at about 500 °C under 50 bar of hydrogen; its structure was solved from neutron powder diffraction of the deuteride in space group $P2_1/c$ with a = 10.226 Å, b = 19.234 Å, c = 8.3345 Å, and it contains one square-pyramidal $[IrH_5]^{4-}$ complex and three saddle-like $[IrH_4]^{5-}$ complexes, five hydride ions being bonded to magnesium only [7].

It is absent from Alexandria PBE, from Alexandria PBEsol and from the OQMD snapshot; we found no Materials Project record for it either. So is $Mg_2IrH_9$, a composition named as a near-hull competitor at 20 GPa in one of the studies [3].

Two consequences follow, and we state both.

1. **Every published hull of this system is incomplete in the same way, ours included.** This is not a criticism of any study; it is a property of the available reference data, and it is checkable in minutes once one thinks to look.
2. **Our verdict is a bound, not a final value.** Adding a phase to a hull can only raise, or leave unchanged, the distance of any compound above it. Including the missing phases can therefore only move $Mg_2IrH_6$ further above the hull. The reported +85.5 meV/atom is a **lower bound**, and the direction of the conclusion is protected by the monotonicity of the construction even though its magnitude is not final.

Computing these two phases is the obvious next step. We began on the first of them; the attempt is reported in Supplementary Note 7.7, and its registered outcome is inconclusive.

The monotonicity argument can be tested rather than cited. No hypothetical phase, at any composition on a grid across the ternary triangle and any formation energy down to −5 eV/atom, makes $Mg_2IrH_6$ stable against the complete set as the theorem requires. Against the reduced set the picture inverts: at **all 66** grid compositions some undiscovered phase would overturn the stable verdict (Supplementary Note 5.3).

One absent compound, however, is an anecdote. The next three subsections ask whether it is a symptom.

## 5.4 The omission is not isolated: a cell-size ceiling in the database that supplies our energies

One synthesised compound absent from every database we queried is an anecdote. Whether it is a property of the reference data is a different question, and it is measurable. We pre-registered the measurement, with one feature that the rest of this paper does not need: because the hypothesis was generated by the two systems already in front of us, those two are a **discovery set** and are excluded from the verdict, which rests on a held-out test set.

The selection rule is mechanical: magnesium, hydrogen, and every transition metal of the three series 29 ternary systems, none dropped for its result, an empty system reported empty and counted. Mg–Ir–H and Mg–Co–H are the discovery set; the other 27 are the test set. The registered criterion was that the test set show a median below 20 atoms per cell and fewer than 1% of entries at 100 atoms or more; 5% or more at 100 atoms would reject.

The criterion is met and the ceiling is confirmed, and it is a statement about **ternary** entries, which is what the query selects and what a ternary hull is built from: across the 4412 ternary test-set entries the median cell holds 10 atoms, the largest holds 81, and **none reaches 100**. The registered criterion was the median and the fraction at 100 atoms, and both are met; everything that follows the repeated integer, the prototype families was noticed afterwards and is post-registered in the sense Supplementary Note 2.6 defines. The restriction to ternary entries is not a technicality.

The same snapshot holds a 152-atom cell in the Mg–Ir–H system itself the binary IrMg, space group 64 along with binaries at 150, 102 and 99 atoms. The generator can evidently reach 152 atoms; what it does not do is produce a **ternary** prototype above 80. What the summary understates is how sharp the limit is. Taking the largest ternary cell in each of the 29 systems separately, **23 stops at exactly 80 atoms**, one at 81 and the remaining five below 34. Natural scarcity would scatter those maxima; a single repeated integer is the signature of a generation limit, and every 80-atom maximum across the 29 systems belongs to one of just **four** stoichiometry families, substituted across the periodic table; in Mg–Ir–H the single one present is $Mg_{12}Ir_3H_5$ (Supplementary Note 5.4).

The ceiling is a description of the data, not an explanation. That it comes from prototype substitution is a reasonable reading of the repeated prototype, not something we can test from outside, since the generation pipelines are not published in that detail.

## 5.6 A structural test, and what it implies for a hull

Two tests have now failed presence by composition, and the ceiling read as a global bound on cell size (Supplementary Note 5.5) and both fail for the same reason: neither asks the question a hull asks. A hull is not built from compositions, and it is not built from a global bound on cell size either. It is built from the energies of **structures**, one per composition. The question that bears on a hull is therefore neither whether the composition appears nor whether any measured cell is large in absolute terms, but whether the database holds, at that composition, a cell large enough to be the measured structure at all. We registered that question separately, stating in advance that it was generated by the data and is therefore a confirmatory test on a wider sample rather than an independent discovery.

A compound counts as **covered** when the database holds an entry at its composition with at least its primitive atom count: necessary and not sufficient, so measured coverage is an upper bound on true coverage and that error runs against the hypothesis. Two other choices run the *other* way and we name them rather than leaving them in the code a compound enters once with the **largest** primitive count recorded for it, registered in advance, and the count is read from a Pearson symbol, which for a disordered refinement counts sites rather than atoms. Both make "not covered" more likely (Supplementary Note 5.6), and the second needed a repair.

**One repair was needed first.** A unit cell holds an integer number of empirical formula units, so its atom count must be divisible by the formula's. Eleven of the 24 harvested rows are not, which proves their Pearson numbers are not atom counts: a disordered refinement reports sites rather than atoms and the number is too large, while an entry indexed on the metal framework alone makes it too small. We replace each such count by the largest value consistent with the site count and the formula, drop one row for which no such value exists and two that name themselves non-stoichiometric, and keep 21 compounds. The substitute is still an upper bound on the experimental count, so the repair removes part of the bias toward the hypothesis and not all of it. Supplementary Note 5.6 gives the rule, every affected row, and the independent confirmation that the two largest cells hold 152 atoms and not the 168 and 156 harvested.

| | Covered | Not covered |
|---|---|---|
| Compounds, test set | 13 | 3 |
| **Median primitive atoms** | **10** | **28** |
| Compounds, all 29 systems | 14 | 7 |
| Median primitive atoms | 11 | 28 |

On the test set the registered primary test, a one-sided Mann–Whitney on primitive atom count, gives **p = 0.016**, the median of the uncovered group is larger, and the verdict is confirmed. Across all 21 compounds it gives **p = 0.0014**. The registered supporting test, a Fisher exact test at the registered threshold of 24 primitive atoms, gives **p = 0.071** on the test set and **p = 0.025** across all 21. **The supporting test therefore does not reach significance on the test set, and we report that rather than quoting the primary alone.** A threshold test discards the rank information the primary test uses and the sample is small, but the result is weaker than a single p-value would suggest.

Both p-values rose under the repair, from 0.0019 and 0.00052 on the harvested numbers, because the repair removed inflation that had been working in the hypothesis's favour. The direction of the effect did not change and neither did either verdict.

The seven uncovered compounds are listed with their measured and stored cell sizes in Supplementary Note 5.6. Three are absent from the database at their composition altogether, including both 152-atom phases; the remaining four are stored at between a quarter and a half of the size they need.

The consequence for a hull follows directly: where the stored structure is too small to be the measured one, its energy is not that compound's ground-state energy but a higher one, the hull there is **too shallow**, and every distance measured against it is a lower bound. Section 5.3 reached the same conclusion from monotonicity for one missing compound; here it has a measured mechanism and a wider base.

The limits are heavy. The funnel runs from 146 harvested records to 93 stoichiometric ternaries, to 24 distinct compounds, to 21 after the repair, and every statistic above is conditioned on it (Supplementary Note 5.6). **Two limits matter more than the denominator and we state them first.** Of the seven uncovered compounds four are in the discovery set and one of the remaining three, $Mg_6VH_{14}$, is absent at its composition altogether the mode Supplementary Note 5.5 found *not* size-dependent so **the genuinely new held-out evidence is two compounds: $Mg_3CrH_8$, 48 primitive atoms against a stored 12, and $Mg_3RuH_3$, 28 against 14**. And "not covered" is *defined* as the measured count exceeding the stored maximum, while those maxima barely vary (9, 10, 12, 14), so the rank test is partly a test of its own definition and could not easily have come out the other way which is the honest reason the supporting test is the weaker of the two. Supplementary Note 5.6 gives both in full, and the test that would separate them, which we have not run.

The remaining limits are as before: coverage is a necessary condition only; one database and one chemical family are surveyed; the atom counts come from a secondary source's Pearson symbols rather than the primary refinements, and eleven of twenty-four needed the repair; and the correlation with size does not establish that generation is its cause.

---

# 7. Recomputing the central number from first principles

## 7.5 The verdict

| Quantity | Database energies, registered before submission | Our energies | Difference |
|---|---|---|---|
| **Cubic $Mg_2IrH_6$ above the hull** | **85.49** | **89.75** | **+4.26** |
| C2/m polymorph above the hull | 69.13 | 76.58 | +7.45 |
| Decomposition facet | $Mg_2IrH_5$ 0.4444 + $Mg_2IrH_7$ 0.5556 | identical to four figures | |

The registered outcome agrees: the difference is 4.26 meV/atom against a band of 15, the sign is positive, the facet is unchanged, and no run failed a validity check. The hull section that our own energies produce, and the facet they select, are shown in Fig. 4a.

Three determinations of the same quantity now exist, two of them independent in geometry as well as in energy (Fig. 4b): 85.49 from the database energies, 89.48 from ours on the database geometries averaged over five settings of cutoff and k-point spacing, and 89.75 from ours on our own relaxed geometries. The last two differ by 0.27 meV/atom, so relaxation at this cutoff does not move the verdict, and all three exceed zero by more than five times the acceptance band.

The central result of this paper therefore does not depend on the database being internally consistent: it survives being recomputed.

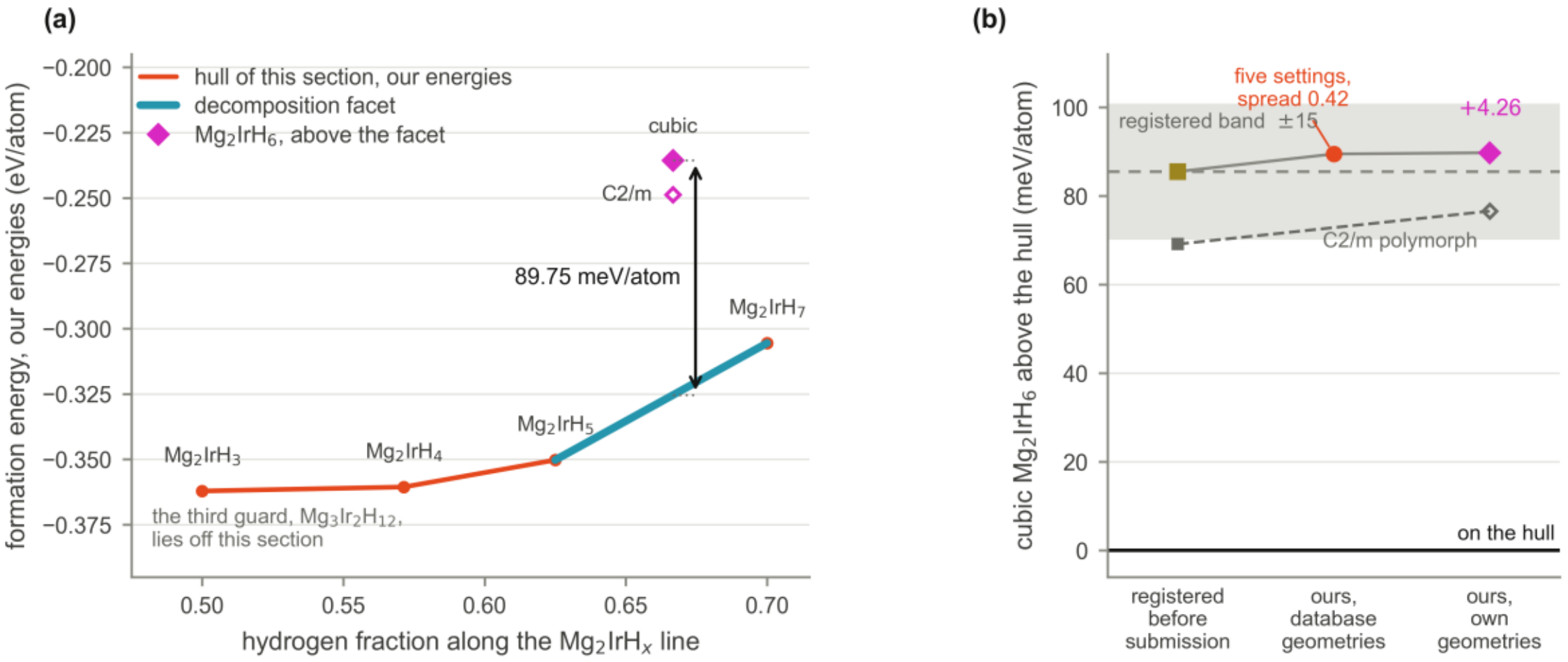

**Figure 4.** (a) The $Mg_2IrH_x$ section of the hull built from our own density-functional energies. The decomposition facet is the $Mg_2IrH_5$–$Mg_2IrH_7$ tie line returned by the frozen comparison script, identical to the one registered before submission, and it leaves cubic $Mg_2IrH_6$ 89.75 meV/atom above the hull; the C2/m polymorph of the same composition lies below the cubic structure and still above the facet. The third guard phase, $Mg_3Ir_2H_{12}$, does not lie on this section and is reported in the data file instead. (b) The registered test. From left to right: the distance registered before any job was submitted, the same quantity from our energies on the database geometries at five settings of cutoff and k-point spacing, and from our own geometries relaxed in two stages. The shaded region is the acceptance band of ±15 meV/atom fixed in advance. The difference is +4.26 meV/atom, the sign is positive and the facet is unchanged, so the registered outcome is agreement.

A second, post-registered derivation sharing no code with the frozen script no linear program and no numerical library, the facet being a tie line whose mixing fraction follows in closed form from the magnesium balance returns 89.75 with the iridium and hydrogen balances closing to numerical zero (Supplementary Note 7.5).

The campaign consumed 20.6 core-hours across 60 completed tasks, no task exceeding one hour of wall time.

## 8. Could a fast universal potential have decided this?

Relaxing 1342 entries with density-functional theory is a campaign; relaxing them with a universal machine-learning potential takes minutes, so if such a potential reproduced the verdict, audits of this kind would be cheap. That is a decision question, and one where a favourable answer is easy to obtain after the fact, so the criteria and the analysis code were fixed before the calculation ran. A 93-structure set was relaxed with MACE-MP-0 (`medium`), and the experiment was then repeated with MACE `medium-0b3` and ORB `orb-mptraj-only-v2`, the two further potentials of the same circularity-free class models trained on broader data have seen our reference set and cannot test agreement with it.

**All three failed, in the same way.** Each places cubic $Mg_2IrH_6$ exactly on its own hull where the reference places it 85.5 meV/atom above, each makes its decomposition product the compound itself rather than the $Mg_2IrH_5$–$Mg_2IrH_7$ tie line, and each misses the registered magnitude limit computed from its own typical error. The failure is therefore not an artefact of one release or one architecture, though the three share a training set and so are one sample rather than three.

**And the error is not a correctable offset**, which is the part that transfers beyond these models. A per-element correction fitted to the model's own error provably moves no hull distance at all this is the invariance of Section 4.2, confirmed numerically at 0.0000 meV/atom while a chemistry-class correction, which can move one, makes the agreement worse under leave-one-out and pushes $Mg_2IrH_6$ further in the wrong direction. Supplementary Notes 8.1 to 8.7 give the registered criteria, the three verdicts with their diagnostics, the calibration fact or with its uncertainty, and the limits of the claim.

## 9. What a reported hull distance must carry

The results above support a short, testable list. A hull distance is interpretable only if it is reported together with:

1. **The reference set**, by source and snapshot date, with the number of entries and of distinct compositions in the relevant chemical system.
2. **The pseudo-hull distance** [12] or an equivalent sensitivity measure, so that fragility is visible.
3. **An add-one sensitivity test** against the nearest absent compositions in this system four compositions out of 212 carried the entire effect, so the test is cheap when it is targeted chemically rather than exhaustively.
4. **The identity of the hydrogen (or other volatile element) reference phase**, including whether a molecular phase is present at all. The defect of Supplementary Note 4.1 is invisible in a formation energy and obvious in one line of metadata.

5. **A statement of which hull phases are experimentally known,** and which are predictions, and in particular whether the decomposition facet of the compound of interest consists of made compounds.
6. **Whether the number is a bound.** Since adding phases can only raise E_hull, a value computed from a database that is known to be missing a synthesised compound is a lower bound and should be labelled one.

Items 1, 2 and 4 are metadata and cost nothing. Items 3, 5 and 6 cost a literature afternoon and a few linear programmes. We provide the protocol that performs them (Supplementary Note B).

---

## 10. Limitations

The full list is Supplementary Note 10, with every item and its evidence. Six bear directly on the result. **Our own reference set is incomplete**: $Mg_6Ir_2H_{11}$ and $Mg_2IrH_9$ are absent from every snapshot we queried, monotonicity protects the direction of the verdict but not its magnitude, and the campaign that would fix the magnitude is unfinished and registered inconclusive. **Our own calculations cover ten phases of 1342**, which the margin analysis of Section 3.4 licenses by bounding the exposure rather than removing it. **The machine-learning experiment covers one training set**, because models trained on broader data have seen our reference data and cannot test agreement with it. **The energetics are static and at 0 K**, and the two finite-temperature terms that might matter are bounded rather than conceded. The facet conserves hydrogen, so the zero-point term of the hydrogen reference cancels exactly and erasing the verdict would need the mean hydrogen frequency of $Mg_2IrH_6$ to lie 724 $cm^{-1}$ below the facet average; we have not computed the residual. The configurational entropy that reference [3] invokes belongs to $Mg_2IrH_5$, which is a decomposition *product*, so at its ideal ceiling it **raises** the distance by 3.9 meV/atom at 300 K and 12.9 at 1000 K and would need about 6900 K to erase it. **The attribution to one phase is read off a published text, not off a rebuilt hull**, so three gaps remain: a structure computed and found metastable leaves no trace in that text, its hull is PBEsol where ours is PBE, and our baseline is not its reference set. **And one chemical system is studied**: the protocol is general; the numbers are not.

## 11. Conclusion

A published disagreement about whether a predicted ambient-pressure superconductor is thermodynamically stable is attributable, within one functional and one protocol, to a single competing phase, worth 131 meV/atom and the change of sign. The identification survives being checked against the published study rather than against our own baseline: of the four compositions that can flip the verdict, two sit on that study's own reported hull while the formula of the decisive fourth occurs nowhere in it. The attribution is asymmetric a stable verdict from an incomplete set is overturned by one phase, an unstable verdict from a complete one survives the removal of both members of its facet  and the other candidate explanation, the hydrogen reference, is a real defect of one data set that is provably incapable of moving any hull distance.

The hull can then be tested against something other than another calculation. The four synthesised hydrides the database contains lie within 9.1 meV/atom of it, three of them within 0.4, the facet that decomposes $Mg_2IrH_6$ consists of two compounds that have both been made, and the compound never made is at least 7.6 times further. A ternary hydride synthesised in 2002 is absent from all three databases we queried for it and pursued across 29 systems that omission has a measurable shape though not the one we first looked for. The reference data hold no ternary entry above 81 atoms per cell and 23 of the 29 systems stop at exactly 80 while the same snapshot carries binary cells to 152 atoms, so the limit is on ternary generation and not on cell size yet no held-out system holds a measured structure above that ceiling, so the ceiling as a global bound excludes nothing we can demonstrate. What survives on held-out data is the per-composition claim, which is what a hull actually uses: the compounds the data cannot accommodate at their own composition are systematically the larger ones, so the hull there is too shallow and every published hull of this system is a bound rather than a value, ours included.

None of this rests on trusting the source. Recomputing the ten decisive phases under one setting of our own, against distances and a facet registered before submission, returns 89.75 meV/atom where the registered value is 85.49, with the facet unchanged; had the sign come out negative we had undertaken to publish that instead. Three universal potentials, asked under criteria fixed in advance, could not arbitrate the question: each places the compound on its own hull and makes its decomposition product itself, and the error is absorbable by neither a per-element nor a chemistry-class correction.

The general lesson is modest and operational. A hull distance is a two-part statement, and only one part is usually reported. Reporting the other part costs almost nothing, and in this system, it is the part that decides the answer.

---

## References

[1] A. Sanna, T. F. T. Cerqueira, Y.-W. Fang, I. Errea, A. Ludwig and M. A. L. Marques, Prediction of ambient pressure conventional superconductivity above 80 K in hydride compounds, *npj Computational Materials* **10**, 44 (2024), DOI 10.1038/s41524-024-01214-9; preprint arXiv:2310.06804, whose title carries the additional words "in Thermodynamically Stable Hydride Compounds".

[2] K. Dolui, L. J. Conway, C. Heil, T. A. Strobel, R. Prasankumar and C. J. Pickard, Feasible route to high-temperature ambient-pressure hydride superconductivity, *Phys. Rev. Lett.* **132**, 166001 (2024), DOI 10.1103/PhysRevLett.132.166001; preprint arXiv:2310.07562.

[3] M. F. Hansen, L. J. Conway, K. Dolui, C. Heil, C. J. Pickard, A. Pakhomova, M. Mezouar, M. Kunz, R. P. Prasankumar and T. A. Strobel, Synthesis of $Mg_2IrH_5$: a potential pathway to

high-T_c hydride superconductivity at ambient pressure, *Phys. Rev. B* **110**, 214513 (2024), DOI 10.1103/PhysRevB.110.214513; preprint arXiv:2406.09538. The synthesis is reported as readily accessible at mild pressure–temperature conditions, and the stability comparison against $Mg_2IrH_6$ covers conditions to about 28 GPa.

[4] S. Sinha, W. Lu, M. F. Hansen, M. J. Hutcheon, T. W. Bontke, L. J. Conway, K. Dolui, C. J. Pickard, C. Heil, P. A. Guńka, S. Chariton, V. Prakapenka, L. Deng, C.-W. Chu, M. N. Julian, R. P. Prasankumar and T. A. Strobel, High-pressure stabilization of $Mg_2IrH_7$: structural proximity to high-T_c superconductivity, arXiv:2602.23675 (submitted 27 February 2026). No journal version was listed on the preprint record when it was checked on 2026-11-05.

[5] L. Wu, Z. Wang, G. Liu, J. Zhang, Y. Ge, Y. Su, R. Chen, H. Liu, W. Li, S. Zhang, J. Zhu, J. Zhao, Z. Deng, S. Feng, J. Song, Q. Liu, X. Li, H. Liu, P. Kong, X. Wang and C. Jin, Superconducting hydride $Mg_2RhH_6$ experimentally achieved at lower pressure, *Journal of the American Chemical Society* **148**, 36854 (2026), DOI 10.1021/jacs.6c08889; preprint arXiv:2608.15745.

[6] W. Lu, M. J. Hutcheon, M. F. Hansen, K. Dolui, S. Sinha, M. R. Sahoo, C. J. Pickard, C. Heil, A. Pakhomova, M. Mezouar, D. Daisenberger, S. Chariton, V. Prakapenka, M. N. Julian, R. P. Prasankumar and T. A. Strobel, Prediction and synthesis of $Mg_4Pt_3H_6$: a metallic complex transition metal hydride stabilized at ambient pressure, arXiv:2505.22546 (submitted 28 May 2025); synthesised in laser-heated diamond anvil cells at 8–25 GPa and recovered to ambient conditions. No journal version was listed on 2026-11-05.

[7] R. Černý, J.-M. Joubert, H. Kohlmann, K. Yvon, *J. Alloys Compd.* **340**, 180–188 (2002). $Mg_6Ir_2H_{11}$, a new metal hydride containing saddle-like $[IrH_4]^{5-}$ and square-pyramidal $[IrH_5]^{4-}$ hydrido complexes. DOI 10.1016/S0925-8388(02)00050-6.

[8] Matbench Discovery: a framework to evaluate machine-learning crystal-stability predictions; *Nature Machine Intelligence* (2025), DOI 10.1038/s42256-025-01055-1; live leaderboard at matbench-discovery.materialsproject.org. Leaderboard values quoted here were read on 2026-10-08.

[9] The Alexandria database of density-functional calculations. PBE set: J. Schmidt, N. Hoffmann, H.-C. Wang, P. Borlido, P. J. M. A. Carriço, T. F. T. Cerqueira, S. Botti and M. A. L. Marques, Machine-learning-assisted determination of the global zero-temperature phase diagram of materials, *Advanced Materials* **35**, 2210788 (2023), DOI 10.1002/adma.202210788. PBEsol and SCAN sets: J. Schmidt, H.-C. Wang, T. F. T. Cerqueira, S. Botti and M. A. L. Marques, *Scientific Data* **9**, 64 (2022), DOI 10.1038/s41597-022-01177-w. We take the `pbe` endpoint to serve the first and the `pbesol` endpoint the second, which is what the functional names imply; the OPTIMADE `info` response carries no citation field, so the mapping is inferred. `[VERIFY with the provider that the served PBE set is the one of the 2023 paper and not a later expansion of it]`

[10] J. E. Saal, S. Kirklin, M. Aykol, B. Meredig and C. Wolverton, Materials design and discovery with high-throughput density functional theory: the Open Quantum Materials Database, *JOM* **65**, 1501–1509 (2013);

and S. Kirklin, J. E. Saal, B. Meredig, A. Thompson, J. W. Doak, M. Aykol, S. Rühl and C. Wolverton, The Open Quantum Materials Database (OQMD): assessing the accuracy of DFT formation energies, *npj Computational Materials* **1**, 15010 (2015), DOI 10.1038/npjcompumats.2015.10.

[11] I. Batatia *et al.*, A foundation model for atomistic materials chemistry, *The Journal of Chemical Physics* **163**, 184110 (2025), DOI 10.1063/5.0297006; preprint arXiv:2401.00096 (2023), which is the version whose release we used. `[VERIFY that the published version describes the same medium release of MACE-MP-0, since it documents several model versions]`

[12] The AFLOW ecosystem of databases, software and tools, including the pseudo-hull distance; arXiv:2207.09842.

[13] Y. An, S. Zhu, J. Montoya, X. Guo and Z. Wang, Predicting novel stable materials for experimental synthesis, arXiv:2607.01713 (submitted 2 July 2026).

[14] Systematic softening in universal machine-learning interatomic potentials; *npj Computational Materials* (2024), DOI 10.1038/s41524-024-01500-6.

[15] Benchmarking universal machine-learning interatomic potentials on elemental systems; *npj Computational Materials* (2026), DOI 10.1038/s41524-026-02251-2.

[16] The Open Materials 2024 (OMat24) inorganic materials dataset and models; *Nature Computational Science* (2026), DOI 10.1038/s43588-026-00996-w; arXiv:2410.12771.

[17] M. A. Alam, M. A. H. Shah, F. Parvin and S. H. Naqib, Physical properties of transition metal hydride superconductors $Mg_2TmH_6$ (Tm = Rh, Pd, Ir, Pt) by first-principles calculations, arXiv:2604.25626 (submitted 28 April 2026).

[18] K. Yvon and G. Renaudin, Hydrides: solid state transition metal complexes, in *Encyclopedia of Inorganic Chemistry*, 2nd edition (John Wiley & Sons, 2006), pp. 1814–1846, DOI 10.1002/0470862106.ia087.

[19] F. Körmann, Y. Ikeda, K. Glazyrin, M. Bykov, K. Spektor, S. Bhat, N. Y. Gugin, A. Bochkarev, Y. Lysogorskiy, B. Grabowski, K. V. Yusenko and R. Drautz, Hydrogen uptake and hydride formation in $Al_x$CoCrFeNi high-entropy alloys: first-principles, universal-potential, and experimental study, arXiv:2603.17479 (submitted 18 March 2026).

[20] C. D. Chambers and L. Tzavella, The past, present and future of Registered Reports, *Nature Human Behaviour* **6**, 29–42 (2022), DOI 10.1038/s41562-021-01193-7

[21] SpringerMaterials, the Inorganic Solid Phases collection derived from the Pauling File, consulted 2026-10-24 for the experimental inventory of Supplementary